\pdfoutput=1 

\documentclass[twoside,twocolumn,9pt]{article}
\usepackage{extsizes}
\usepackage[super,sort&compress,comma]{natbib} 
\usepackage[version=3]{mhchem}
\usepackage[left=1.5cm, right=1.5cm, top=1.785cm, bottom=2.0cm]{geometry}
\usepackage{balance}
\usepackage{mathptmx}
\usepackage{sectsty}
\usepackage{graphicx} 
\usepackage{lastpage}
\usepackage[format=plain,justification=justified,singlelinecheck=false,font={stretch=1.125,small,sf},labelfont=bf,labelsep=space]{caption}
\usepackage{float}
\usepackage{fancyhdr}
\usepackage{fnpos}
\usepackage[english]{babel}
\addto{\captionsenglish}{%
  
}
\usepackage{array}
\usepackage{droidsans}
\usepackage{charter}
\usepackage[T1]{fontenc}
\usepackage[usenames,dvipsnames]{xcolor}
\usepackage{setspace}
\usepackage[compact]{titlesec}
\usepackage{hyperref}

\usepackage{epstopdf}

\definecolor{cream}{RGB}{222,217,201}

\begin{document}

\pagestyle{fancy}
\thispagestyle{plain}
\fancypagestyle{plain}{
\renewcommand{\headrulewidth}{0pt}
}

\makeFNbottom
\makeatletter
\renewcommand\LARGE{\@setfontsize\LARGE{15pt}{17}}
\renewcommand\Large{\@setfontsize\Large{12pt}{14}}
\renewcommand\large{\@setfontsize\large{10pt}{12}}
\renewcommand\footnotesize{\@setfontsize\footnotesize{7pt}{10}}
\makeatother

\renewcommand{\thefootnote}{\fnsymbol{footnote}}
\renewcommand\footnoterule{\vspace*{1pt}%
\color{cream}\hrule width 3.5in height 0.4pt \color{black}\vspace*{5pt}} 
\setcounter{secnumdepth}{5}

\makeatletter 
\renewcommand\@biblabel[1]{#1}            
\renewcommand\@makefntext[1]%
{\noindent\makebox[0pt][r]{\@thefnmark\,}#1}
\makeatother 
\renewcommand{\figurename}{\small{Fig.}~}
\sectionfont{\sffamily\Large}
\subsectionfont{\normalsize}
\subsubsectionfont{\bf}
\setstretch{1.125} 
\setlength{\skip\footins}{0.8cm}
\setlength{\footnotesep}{0.25cm}
\setlength{\jot}{10pt}
\titlespacing*{\section}{0pt}{4pt}{4pt}
\titlespacing*{\subsection}{0pt}{15pt}{1pt}

\fancyfoot{}
\fancyfoot[RO]{\footnotesize{\sffamily{  \hspace{2pt}\thepage }}}
\fancyfoot[LE]{\footnotesize{\sffamily{ \thepage}}}
\fancyhead{}
\renewcommand{\headrulewidth}{0pt} 
\renewcommand{\footrulewidth}{0pt}
\setlength{\arrayrulewidth}{1pt}
\setlength{\columnsep}{6.5mm}
\setlength\bibsep{1pt}

\makeatletter 
\newlength{\figrulesep} 
\setlength{\figrulesep}{0.5\textfloatsep} 

\newcommand{\topfigrule}{\vspace*{-1pt}%
\noindent{\color{cream}\rule[-\figrulesep]{\columnwidth}{1.5pt}} }

\newcommand{\botfigrule}{\vspace*{-2pt}%
\noindent{\color{cream}\rule[\figrulesep]{\columnwidth}{1.5pt}} }

\newcommand{\dblfigrule}{\vspace*{-1pt}%
\noindent{\color{cream}\rule[-\figrulesep]{\textwidth}{1.5pt}} }

\makeatother

\renewcommand{\d}{{\rm d}}
\newcommand{\e}{{\rm e}}
\newcommand{\x}{{\rm x}}
\newcommand{\y}{{\rm y}}
\newcommand{\z}{{\rm z}}
\newcommand{\w}{{\rm w}}
\newcommand{\va}{{\sigma}}
\newcommand{\eq}{\begin{equation}}
\newcommand{\eqend}{\end{equation}}
\newcommand{\eqn}[1]{(\ref{#1})}
\newcommand{\Eqn}[1]{Eq.\ (\ref{#1})}
\newcommand{\PD}[2]{\frac{\partial#1}{\partial#2}}
\newcommand{\DD}[2]{\frac{\d#1}{\d#2}}
\newcommand{\pv}{{\bf p}}
\newcommand{\rv}{{\bf r}}
\newcommand{\qv}{{\bf q}}
\newcommand{\vv}{{\bf v}}
\newcommand{\kv}{{\bf k}}
\newcommand{\mv}{{\bf m}}
\newcommand{\nv}{{\bf n}}
\newcommand{\Fv}{{\bf F}}
\newcommand{\Rv}{{\bf R}}
\newcommand{\Ev}{{\bf E}}
\newcommand{\Jv}{{\bf J}}
\newcommand{\Hv}{{\bf H}}
\newcommand{\Mv}{{\bf M}}
\newcommand{\Dv}{{\bf D}}
\newcommand{\uv}{{\bf u}}
\newcommand{\ev}{{\bf e}}
\newcommand{\sv}{{\bf s}}
\newcommand{\lv}{{\bf l}}
\newcommand{\av}{{\bf a}}
\newcommand{\Lv}{{\bf L}}
\newcommand{\Av}{{\bf A}}
\newcommand{\xv}{{\bf x}}
\newcommand{\Sv}{{\bf S}}
\newcommand{\Bv}{{\bf B}}
\newcommand{\Qv}{{\bf Q}}

\twocolumn[
  \begin{@twocolumnfalse}
\vspace{6em}
\sffamily
\begin{tabular}{m{4.5cm} p{13.5cm} }

& \noindent\LARGE{\textbf{Non-Uniform Magnetic Fields for Single-Electron Control$^\dag$}} \\
\vspace{0.3cm} & \vspace{0.3cm} \\

 & \noindent\large{Mauro Ballicchia,$^{\ast}$\textit{$^{a}$} Clemens Etl,\textit{$^{a}$} Mihail Nedjalkov,\textit{$^{a}$} Josef Weinbub\textit{$^{a}$}}  \vspace*{0.8cm} \\

& \noindent\normalsize{Controlling single-electron states becomes increasingly important due to the wide-ranging advances in electron quantum optics. Single-electron control enables coherent manipulation of individual electrons and the ability to exploit the wave nature of electrons, which offers various opportunities for quantum information processing, sensing, and metrology. 
A unique opportunity offering new degrees of freedom for single-electron control is provided when considering non-uniform magnetic fields. Considering the modeling perspective, conventional electron quantum transport theories are commonly based on gauge-dependent electromagnetic potentials. A direct formulation in terms of intuitive electromagnetic fields is thus not possible. 
In an effort to rectify this, a gauge-invariant formulation of the Wigner equation for general electromagnetic fields has been proposed in [Nedjalkov \textit{et al., Phys. Rev. B.}, 2019, \textbf{99}, 014423]. However, the complexity of this equation requires to derive a more convenient formulation for linear electromagnetic fields [Nedjalkov \textit{et al., Phys. Rev. A.}, 2022, \textbf{106}, 052213]. This formulation directly includes the classical formulation of the Lorentz force and higher-order terms depending on the magnetic field gradient, that are negligible for small variations of the magnetic field. In this work, we generalize this equation in order to include a general, non-uniform electric field and a linear, non-uniform magnetic field. The thus obtained formulation has been applied to investigate the capabilities of a linear, non-uniform magnetic field to control single-electron states in terms of trajectory, interference patterns, and dispersion. This has led to explore a new type of transport inside electronic waveguides based on snake trajectories and also to explore the possibility to split wavepackets to realize edge states.   
} 
\end{tabular}

 \end{@twocolumnfalse} \vspace{0.6cm}

  ]

\renewcommand*\rmdefault{bch}\normalfont\upshape
\rmfamily
\section*{}
\vspace{-1cm}


\footnotetext{\textit{$^{a}$~Institute for Microelectronics, TU Wien, Gusshausstrasse 27-29, 1040 Wien, Austria. E-Mail: mauro.ballicchia@tuwien.ac.at}}




\section{Introduction}

The field of electron quantum optics studies and applies electromagnetic (EM) field-controlled phenomena
for manipulating electron states in solid-state quantum systems~\cite{Hoodbhoy2018, Mondal2018,Zhang2020,Karmakar2021,Hoodbhoy_2021}.  
Ideally, transport theories should describe both, processes governing the electron trajectories as well as wave phenomena such as interference and diffraction in multiple dimensions. 
Theories based on EM potentials rely on the formal mathematical apparatus related to the choice of the scalar and vector potentials, which, however, obscures the physical aspects and thus the heuristic understanding of the electron evolution. This is also true for the Wigner theory which among the alternative formulations of quantum mechanics utilizes classical concepts of a phase space and a quasi-distribution function. Indeed, the  underlying Wigner quantum mechanics is formulated in electrostatic terms under the choice of zero vector potential~\cite{Nedjalkov2011Wigner}.  Central quantities in the derived evolution (transport) equation are the Wigner function $f_w$ and Wigner potential $V_w$, defined by the Weyl transform of the density matrix and the scalar potential, respectively. 
The inclusion of the magnetic field, however, introduces additional terms, which depend on the choice of the gauge~\cite{Houston40, Novakovic2011, Krieger86,Levinson70,Kubo64,Rossi98, Iafrate17,Materdey03n1,Materdey03n2,Bellentani2019,Avazpour2022}. 
Six decades ago Stratonovich~\cite{Stratonovich56} generalized the Weyl transform to replace the canonical momentum as a phase space coordinate  with the kinetic momentum. The latter, being a physical quantity, is gauge invariant, and so is the transport equation for $f_w$ derived by the Weyl-Stratonovich transform. Details can be found, e.g., in~\cite{Serimaa86, PRB, PRA} and the references therein. The potentials are completely removed from the theory, which 
offers the advantage of depending on physical factors only, such as
the  EM fields $\Ev,\Bv$.
The equation is mathematically challenging as it depends on multi-dimensional integrals of $f_w$ with the terms $D^F(\Ev)$, $H^F(\Bv)$, and $I^F(\Bv)$, which include the Fourier transform 
$FT=\int d\sv e^{-\frac i\hbar\pv \sv}$
of  the quantities 
$\int_{-1}^1d\tau\left(\sv\cdot\Ev(  \rv+ \frac{\sv\tau}{2})\right)$
and
$\int_{-1}^1d\tau\left(\sv\times\Bv(  \rv+\frac{\sv\tau}{ 2})\right)$, respectively.
Moreover, contrary to the electrostatic counterpart, which has been analyzed and applied for more than three decades to a plethora of quantum transport processes~\cite{querlioz2013, FerryMixiBook2018}, there is limited experience with the properties of the gauge-invariant equation.
It is thus desirable to gain first experience by reducing the complexity to simplify the equation. A first step in this direction is based on the fact that for stationary EM conditions and a homogeneous $\Bv$ the $I^F$-term can be neglected. Also, $H^F$ introduces the magnetic Lorentz force, while $D^F$ can be expressed via $V_w$ for a stationary $\Ev$~\cite{PRA}. 
As a homogeneous magnetic field can be associated to the zeroth-order term in the Taylor expansion of $\Bv(\rv)$, as a next step, it is reasonable to take into account the next (linear) term in the expansion. Therefore, in what follows, we focus on the effects introduced by a magnetic field with a spatial linearity.
Therefore, we consider the following physical settings: The dimensionality of the problem is reduced to 
a two-dimensional electron evolution in the $\rv=(x,y)$-plane.  The inhomogeneous magnetic field is normal to the plane in $z$-direction: $\Bv={(0,0,B(y))}$ . In this way, the Lorentz magnetic force is in the plane. The $y$-coordinate is chosen along the linearity of $\Bv$, so that  $B(y)=B_0+B_1y$. The electric field $\Ev(\rv)$ has a general spatial dependence. Being stationary, the electric field allows 
to reintroduce 
the Wigner potential $V_w(\pv,\rv)$, a quantity which has been physically well analyzed over the last decades. 
In Section~\ref{Sec_WigEquLinMa}, we formulate the corresponding magnetic field aware evolution equation for the Wigner function. An analysis of the operators, which compose the equation, is presented. 

In section~\ref{tri_sec}, certain  physical effects incorporated in the solution are identified  in the case of a weak non-linearity. Interesting spatial correlations between the electric and magnetic fields are observed, which affect the process of magnetotunneling \cite{Pratley2013,Prasad2021}. They are observed in both, density and negativity distributions obtained from the Wigner solution, suggesting the existence of both local and non-local interplays of these fields, and indicating effects, which can be used for controlling the electron evolution. In particular, different settings of the non-linear magnetic field can be used to guide electron trajectories to a desired region of the space.
Furthermore, it is shown that certain evolution patterns such as snake trajectories  and edge states~\cite{Muller1992,Reijniers2000,Mondal2018}, which are expected from classical considerations, are maintained by the quantum evolution. They can be used to guide and manipulate an electron state by restraining it  in a desired region, splitting the density distribution, or by affecting its spreading.

\section{Wigner Equation for Linear Magnetic Fields}\label{Sec_WigEquLinMa}
The equation has been initially formulated for the case of both linear electric and magnetic fields (see equation (25) in~\cite{PRA}). A linear electric field can be accounted for either by an accelerating (Newtonian) force or, equivalently, by the corresponding Wigner potential term. This duality has been used to verify certain quantum particle concepts used in the electrostatic Wigner theory~\cite{Nedjalkov2013Schwaha}.
For an electric field with a general spatial shape it is straightforward  to reintroduce the Wigner potential in the evolution equation (using equations (44) and (45) from~\cite{PRA}): 
\begin{eqnarray} \label{eq:Wiglinmag}
&&\left(\frac{\partial}{\partial t}
+
\frac{\pv}{m}\cdot
\frac{\partial}{\partial \rv}
+\Fv(B(y))
\cdot\frac{\partial}{\partial \pv}\right)f_w\bigl(\pv,\rv\bigr)
=\nonumber
\\
&& \int d\pv' V_w(\pv-\pv',\rv) 
f_w(\pv',\rv) +
\label{lform}\\ 
&&\frac {B_1\hbar^2}m\frac e{12}
\left(
\frac{\partial^2}{\partial p_y^2}\frac{\partial}{\partial x}
-
\frac{\partial}{\partial p_x}\frac{\partial}{\partial p_y}
\frac{\partial}{\partial y}\right)f_w\bigl(\pv,\rv\bigr) 
\nonumber
\end{eqnarray}
The left-hand side contains the Liouville operator ($Lo(B(y))$), where $\Fv$ is the magnetic Lorentz force $\Fv(B(y))=\frac{e}{m}{\pv\times\Bv(\rv)}=\frac{e}{m} (p_y B(y) , -p_x B(y), 0 )$ 
\footnote{In the Lorentz force, in equation~\ref{eq:Wiglinmag}, and in the following, $\pv$ refers to the kinetic momentum instead of the canonical momentum.}
.
On the right-hand side is the Wigner potential term, followed by a term,  which involves higher-order mixed derivatives. 
Without EM fields the equation reduces to $Lo(0)=0$, which resembles the force-less Vlasov equation.
The equation involves Newtonian trajectories, however, the quantum character of the evolution depends on the initial condition~\cite{Dias2004admissible}: The latter can contain, e.g., negative values in contrast to the classical distribution function. An electric field is accounted for by the Wigner potential term ${\cal V}_w$, so that the equation takes the well-known 
form $Lo(0)={\cal V}_w$. $Lo(0)$ is associated with force-less 
Newtonian trajectories, however, this does not challenge the quantum character of the theory: $Lo(0)$ together with ${\cal V}_w$  gives rise to interference, non-locality, tunneling, negativity, and oscillatory behavior of $f_w$.
The evolution is fully coherent as the theory is fully equivalent to wave mechanics~\cite{Dias2004admissible,tatarskiui1983wigner}. 
Next, if $B_0\ne 0$, \eqn{lform} becomes the homogeneous magnetic field equation discussed in~\cite{PRB} (equation (49)). 
The Liouville operator in equation \eqn{lform} involves Newtonian trajectories driven by the inhomogeneous magnetic field $B(y)=B_0+B_1y$, which affects the interplay with ${\cal V}_w$. The last operator with the higher-order derivatives is proportional to $B_1$. This fact provides the opportunity to discriminate the two operators by considering a small $B_1$. 
For large $y$, the magnitude of the linear component $B_1y$ can become larger than $B_0$ so that if both have opposite signs, the sign of the magnetic field $B(y)$ is changed. 
Furthermore, an analysis of physically relevant settings shows that the higher-order derivatives operator can be a few orders of magnitude smaller than the Wigner potential. The latter is characterized by the quantity $\gamma\simeq 10^{14-15}/sec$~\cite{Nedjalkov2011Wigner}, characterizing the electric conditions in nanostructures. This is a quantity equivalent to the total out-scattering rate in Boltzmann transport models, which, for instance for phonons is $10^{12-13}/sec$. These considerations suggest that the interplay of $Lo(B(y))$ and $V_w$ can give rise to important physical effects which dominate the transport for small $B_1$ values. 
Accordingly, we neglect the last row in \eqn{lform} and consider the equation $Lo(B(y))={\cal V}_w$. It resembles the standard Wigner equation, which is associated with a signed-particle model~\cite{Nedjalkov2011Wigner}\footnote{The inclusion of the higher-order derivative terms demands an advanced quantum particle model, which is currently in development~\cite{Clemens}.}. 
The signed-particle model represents an electron state by an ensemble of numerical particles. The particles have special features that carry the quantum information, however, they evolve in the phase space over Newtonian trajectories dictated by $Lo$. Thus
the left-hand side of \eqn{lform} determines the trajectories of the particles in the evolving ensemble  so that the effects of the classical Lorentz force (due to the linear magnetic field) are incorporated also in the quantum evolution. This is demonstrated by the existence of snake and edge modes in the quantum evolution considered in the next section: There, we consider the essential cases of magnetotunneling and  ways to manipulate the state evolution by choosing different settings for $B_0$ and $B_1$.
Finally, it is important to highlight that such level of physical insight is a unique feature of the Wigner formalism.



\section{\label{tri_sec}Magnetic Field Effects in the Quantum Evolution}

In general, a magnetic field is able to control the trajectory of a classical charged particle. However, the quantum case has a wave-like nature which needs to be incorporated. 
This is possible by using a signed-particle model which is a numerical model to describe the evolution of a quantum electron state and, moreover, the magnetic field effect in this evolution. In the following sub-sections, we show and study the effect of applying a non-uniform linear magnetic field to a selection of specific quantum electron state evolution scenarios, and we highlight the use of magnetic fields for controlling the electron state.  Where possible, we compare to classical transport results, as the Wigner signed-particle model we employ provides a seamless transition to the classical transport picture. In Section~\ref{sec_magnetotunneling}, we analyze a magnetotunneling structure where the electron state interacts with an "electric" potential barrier and an external magnetic field orthogonal to the trajectory. In Section~\ref{sec_snake_edge}, we analyze the evolution of a quantum electron state in an electronic waveguide with two opposite external magnetic field configurations, resulting in a so-called snake trajectory and the formation of an edge state, respectively.

\subsection{Magnetotunneling} \label{sec_magnetotunneling}

Magnetotunneling refers to the effects of a magnetic field on tunneling processes of an electron state through a single-barrier potential. It has been analyzed in the framework of the Wigner formalism in~\cite{Kluksdahl89Ferry}, showing how a coherence pattern remains after a barrier, revealing a sort of "two-peak" waveform. In other work, magnetotunneling has been investigated in~\cite{PRB} with a comparison between the classical and the quantum behaviour in presence of a uniform magnetic field $B(y)=B_0$. The "classical" behaviour corresponds to treating the potential barrier as classical electric force. 
In this case, the particle, having a kinetic energy less than the barrier, is completely reflected.
In the quantum case it was demonstrated that there is tunneling, as was expected, e.g., from~\cite{Kluksdahl89Ferry}, but in addition it was shown that the process of transmission (including the interference effects) is clearly affected by the magnetic field. 
In the following, we are going to deepen this analysis, in particular, studying the case where a non-uniform magnetic field is included in the Liouville operator, i.e., where $Lo(B(y))$ includes a non-uniform magnetic field $B(y)=B_0+B_1 y$, and the barrier is quantum mechanically represented by the Wigner potential ${\cal V}_w$.    


\begin{figure*}[htbp]
  \begin{minipage}[t]{0.48\textwidth}
    \centering
    \renewcommand{\thefigure}{1}  
    \includegraphics[width=0.6\linewidth]{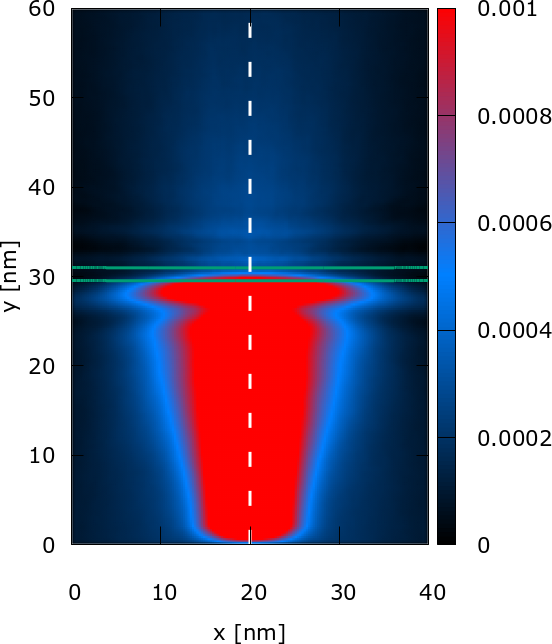}
    \put(-3,26){\hskip-5.8cm \scriptsize{0T --}}
    \put(-3,63){\hskip-5.8cm \scriptsize{0T --}}
    \put(-3,100){\hskip-5.8cm \scriptsize{0T --}}
    \put(-3,137){\hskip-5.8cm \scriptsize{0T --}}
    \put(-3,174){\hskip-5.8cm \scriptsize{B(y)}}
    \vskip -.3cm
    \caption{Case 1: The steady-state electron density, $n(x,y)$, of a Wigner state (electron) injected at the bottom, evolving towards $+y$-direction. No magnetic field is applied (see $B(y)$ indicators on the left). 
    The dashed line indicates the mean path of the state's evolution. Green lines indicate the $1\,\rm nm$ thick barrier of $0.3\,\rm eV$. The density shows a fine oscillatory structure above the barrier.}
    \label{edno}
  \end{minipage}%
  \hfill
  \begin{minipage}[t]{0.48\textwidth}
    \centering  
    \renewcommand{\thefigure}{2}
    \includegraphics[width=0.6\textwidth]{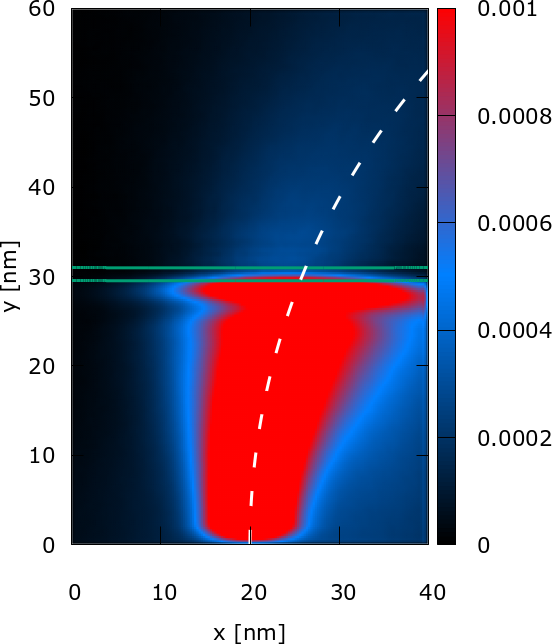}
    \put(-3,26){\hskip-5.8cm \scriptsize{-6T --}}
    \put(-3,63){\hskip-5.8cm \scriptsize{-6T --}}
    \put(-3,100){\hskip-5.8cm \scriptsize{-6T --}}
    \put(-3,137){\hskip-5.8cm \scriptsize{-6T --}}
    \put(-3,174){\hskip-5.8cm \scriptsize{B(y)}}
    \vskip -.3cm
    \caption{Case 2: The steady-state electron density, $n(x,y)$, of a Wigner state (electron) injected at the bottom under a constant magnetic field (see $B(y)$ indicators on the left)  which bends the density
    and thus the mean path, indicated by the dashed line. The oscillatory structure above the barrier is strongly reduced by the magnetic field. }
    \label{dwe}
  \end{minipage}\vspace{0.2cm}
\end{figure*}

\subsubsection{Simulation Setup}

\begin{figure*}[htbp]  
  \begin{minipage}[t]{0.48\textwidth}
    \centering  
    \renewcommand{\thefigure}{3}
    \includegraphics[width=0.6\textwidth]{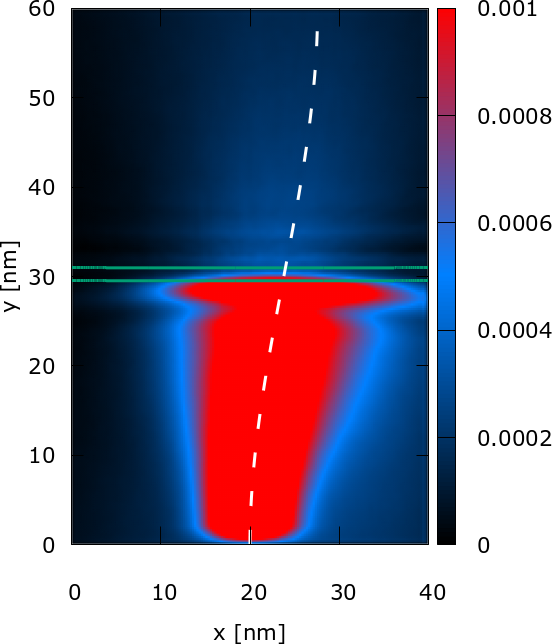}
    \put(-3,26){\hskip-5.8cm \scriptsize{-6T --}}
    \put(-3,63){\hskip-5.8cm \scriptsize{-3T --}}
    \put(-3,100){\hskip-5.8cm \scriptsize{0T --}}
    \put(-3,137){\hskip-5.8cm \scriptsize{3T --}}
    \put(-3,174){\hskip-5.8cm \scriptsize{B(y)}}
    \vskip -.3cm
    \caption{Case 3: The steady-state electron density, $n(x,y)$, of a Wigner state (electron) injected at the bottom, under a linear magnetic that goes from $-6\,\rm T$ to  $6\,\rm T$ (see $B(y)$ indicators on the left) and thus the magnetic field becomes zero at the barrier and switches the sign, giving rise to a snake type of evolution. The density in the upper half of the domain shows again a fine oscillatory structure.}
    \label{tri}
  \end{minipage}%
  \hfill
  \begin{minipage}[t]{0.48\textwidth}
    \centering  
    \renewcommand{\thefigure}{4}
    \includegraphics[width=0.6\textwidth]{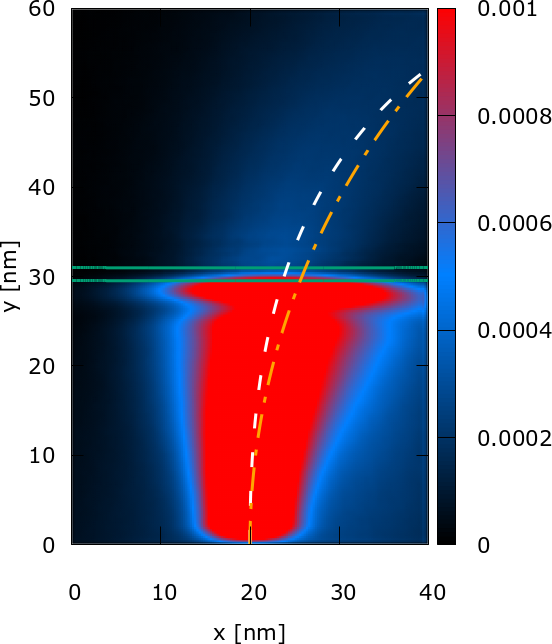}
    \put(-3,26){\hskip-5.9cm \scriptsize{-2\,T --}}
    \put(-3,63){\hskip-5.9cm \scriptsize{-5\,T --}}
    \put(-3,100){\hskip-5.9cm \scriptsize{-8\,T --}}
    \put(-3,137){\hskip-6.0cm \scriptsize{-11\,T --}}
    \put(-3,174){\hskip-5.9cm \scriptsize{B(y)}}
    \vskip -.3cm
    \caption{Case 4: The steady-state electron density $n(x,y)$, of a Wigner state (electron) is injected at the bottom. The magnetic field is gradually increased towards $+y$-direction and is particularly large at and above the barrier (see $B(y)$ indicators on the left). The magnetic field suppresses the oscillations of the density, similar to case 2. The mean path (white dashed line) is compared to the mean path of case 2 (orange dot-dashed line): Although they differ, they both guide the state to the same final position.}
    \label{four}
  \end{minipage}
\end{figure*}

The simulation domain is $(x, y)=(40\, \rm nm, 60\, \rm nm)$ as shown in Figs.~\ref{edno}-\ref{four}. 
A $0.3\,\rm eV$ potential barrier is placed at $y=30 \, \rm nm$. The barrier thickness is $1\, \rm nm$ and is physically modeled by the Wigner potential $V_w$ in all shown experiments. 
The initial condition corresponds to a minimum uncertainty Wigner state $\sigma_{x} = \sigma_{y}=3\, \rm nm$, is periodically injected into the simulation domain from the bottom, and evolves towards the barrier with a kinetic energy of $0.1\, \rm eV$. The initial mean velocity is $0$ in $x$-direction so almost all the energy at $t=0\, \rm fs$ is directed towards the potential barrier. We consider four cases of $(B_0,B_1)$:

\begin{itemize}

\item[] \underline{Case 1} is $(B_0,B_1) = (0,0)$ and represents a reference case for the entire experiment since it allows to analyse quantum tunneling through the barrier without the influence of an external magnetic field. 

\item[] \underline{Case 2} is $(B_0,B_1) = (-6 \,\textrm{T},0)$ and represents the effect of a uniform magnetic field directed orthogonal to the simulation domain.

\item[] \underline{Case 3} is $(B_0,B_1) = (-6\, \textrm{T},0.2\,\textrm{T}\textrm{nm}^{-1})$. 
The magnetic field changes its sign around the barrier; the magnetic field is thus zero at the barrier. 

\item[] \underline{Case 4} is $(B_0,B_1) = (-2\,\textrm{T}, -0.2\,\textrm{T}\textrm{nm}^{-1})$. 
The linear component $B_1 y$ increases the effect of the magnetic field $\Bv$ along $y$ so that the magnitude of the magnetic field is non-zero in the region of the barrier. 

\end{itemize}

\begin{figure*}[htbp]
  \begin{minipage}[t]{0.48\textwidth}
   \centering  
    \renewcommand{\thefigure}{5}
    \includegraphics[width=0.9\textwidth]{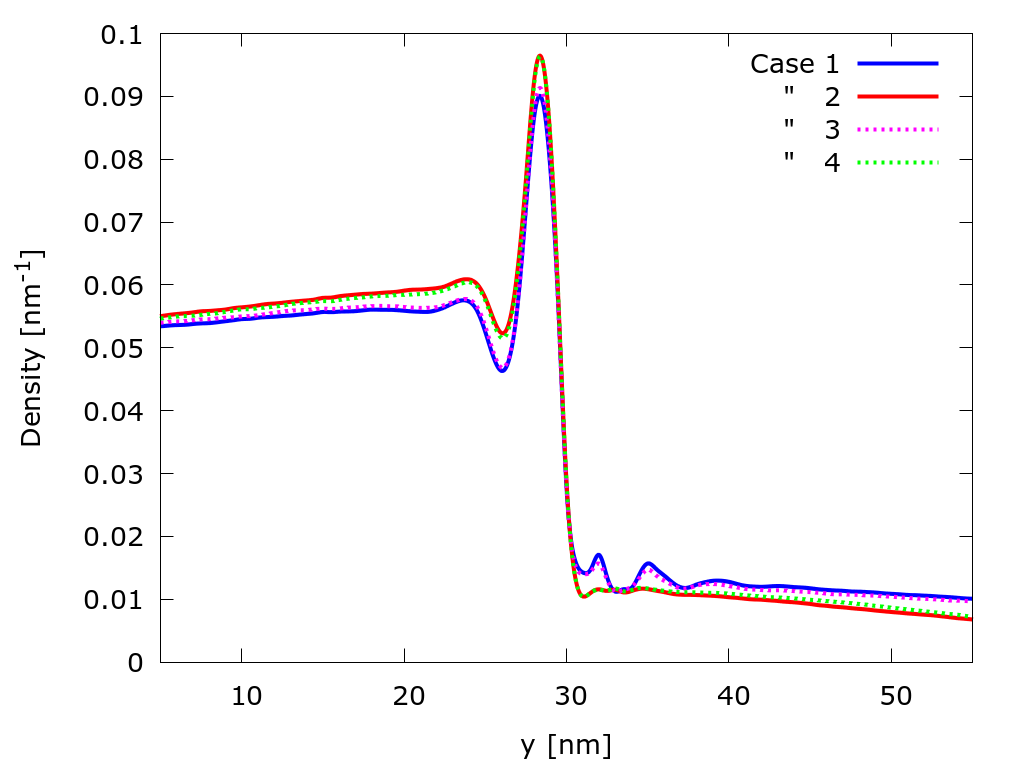}
    \vskip -.3cm
    \caption{Density distribution along $y$-direction $n(y)$:
    Cases 1\&3 and 2\&4 clearly group together, suggesting that the oscillations are suppressed in the
    presence of a magnetic field in the region of the barrier, further indicating that the EM fields interact locally.
    }
    \label{pet}
  \end{minipage}\vspace{0.4cm}
  \hfill
  \begin{minipage}[t]{0.48\textwidth}
    \centering  
    \renewcommand{\thefigure}{6}
    \includegraphics[width=0.9\textwidth]{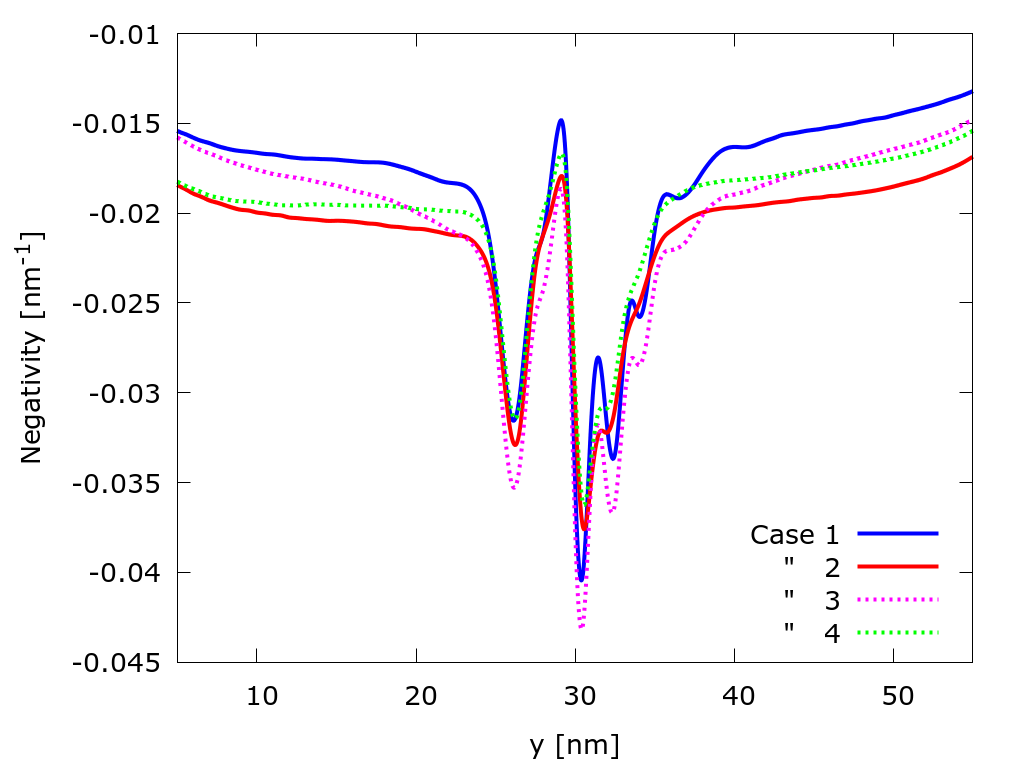}
    \vskip -.3cm
    \caption{Negativity along $y$-direction $Neg_w(y)$. Negative values after the injection ($y<25\,\rm nm$) demonstrates the non-local action of the barrier already without magnetic field. The negativity increases with the increase of $B(y)$ in this region, which suggests again a non-local interplay of the EM fields. Oscillatory behavior is slightly visible for the first two peaks in presence of strong magnetic field: case 2\&4.
    }
    \label{shes}
  \end{minipage}\vspace{0.4cm}
\end{figure*}

\subsubsection{Density and Negativity Analysis}

Figs.~\ref{edno}-\ref{four} shows the electron density for all the four cases.
The steady-state electron density, $n(x,y)$, is obtained by the integration over $\pv$ of the steady-state Wigner function $f_w(\xv,\pv)$. 
In accordance with the Ehrenfest theorem, in all the four cases the mean densities follow the classical paths, that are indicated by the dotted lines. 

In Fig.~\ref{edno}, that is related to $B(y)=0$, the classical path is the central line $x=20\, \rm nm$, since no magnetic field bends the trajectory. The density perfectly reflects the symmetry with respect to the central dashed line and shows a fine oscillatory structure above the barrier, after the tunneling.
The latter is almost completely destroyed by the constant magnetic field, as shown in Fig.~\ref{dwe}.
There, the magnetic field bends the path, guiding the electron towards a specific position, where, considering possible applications, an additional channel could be envisioned, outlining a possible use-case for single electron control.

In case 3 the magnetic field changes its sign around the barrier, giving rise to a snake type of evolution~\cite{Hoodbhoy2018}, as shown in Fig.~\ref{tri}. 
Besides, the fine structure of the density above $y=30nm$ is recovered, similar to the case shown in Fig.~\ref{edno}. 
Observing that the magnetic field is zero at the barrier as in Fig.~\ref{edno}, we associate this effect with the existence of a local interplay with the EM fields. 
Indeed, in Fig.~\ref{four}, when $B(y)$ around the barrier is particularly large (similar to case 2), the oscillations are again suppressed. 

In case 4  the electron density, shown in Fig.~\ref{four}, is very similar to the electron density of case 2, since the magnetic field is negative in the entire simulation domain, but gradually increases towards $+y$-direction starting from $-2\, \rm T$ until $-15\, \rm T$. The mean path (white dashed line) is compared to the mean path of case 2 (orange dot-dashed line): Although they differ, they both guide the state to the same final position. 
An important observation is that the magnetic field is particularly large at and above the barrier in this case, and indeed it suppresses the oscillations of the density, similar to case 2 (Fig.~\ref{dwe}).

Since the linear change of the magnetic field is along the $y$ axis and no significant asymmetries with respect to the classical trajectory are observed along $x$, to better analyze the four cases, the steady-state electron density $n(x,y)$ has been integrated over $x$, obtaining the density along $y$, i.e., $n(y)$.  
One of the advantages of the Wigner formalism and, in particular, of the signed-particle model is to have access to Wigner function and its negativity, obtained by $Neg_w(\rv,\pv)= f_w \theta(-f_w)$ where $\theta$ is the Heaviside function. Wigner function negativity indicates quantum behavior of a state~\cite{Ballicchia2019InvestigatingQC}.
We thus also investigate the negativity along $y$, $Neg_w(y)$, integrating over the momentum $\pv$ and over $x$.

From the comparison of $n(y)$ shown in Fig.~\ref{pet} it is possible to notice that the oscillatory behaviour in case 1 is smaller than in case 2 before the barrier. It is almost  completely destroyed by the magnetic field after the barrier in case 2, only a small variation related to the first two peaks is slightly visible, while in case 1 it is well visible. A further confirmation of the action of the magnetic field on the oscillatory behaviour is given by the fact that case 3, where the magnetic field is almost zero around the potential barrier, behaves similar to case 1, while in case 4, where the magnetic field is even stronger than $-6\, \rm T$, the behaviour is very similar to case 2. 
This observation strongly suggests the existence of a local effect of the magnetic field on the process of tunneling. 

The negativity, shown in Fig.~\ref{shes}, demonstrates an oscillatory behaviour around the barrier, which is a manifestation of quantum effects. 
The appearance of negative values after the injection of the entirely positive initial state below the barrier ($y<25\,\rm nm$) demonstrates the non-local action of the barrier already without a magnetic field.
The oscillations of the negativity are well visible after the barrier for case 1 and 3, while they are drastically reduced for case 2 and 4: Only the first two peaks are slightly visible as a change of the slope (gradient). In any case the negativity shows this behaviour more evidently than the density. This observation provides a confirmation that the magnetic field tends to destroy the oscillatory behaviour of tunneled electrons and also confirms the results presented in~\cite{Kluksdahl89Ferry}. It's also possible to observe that the negativity increases with the magnitude of the magnetic field already far before the barrier, which can be seen in all the cases where a magnetic field is different from zero, i.e., cases 2,3, and 4. This suggests another, non-local effect of the interplay of the EM fields. 





\begin{figure}[htbp]
\includegraphics[width=0.45\textwidth]{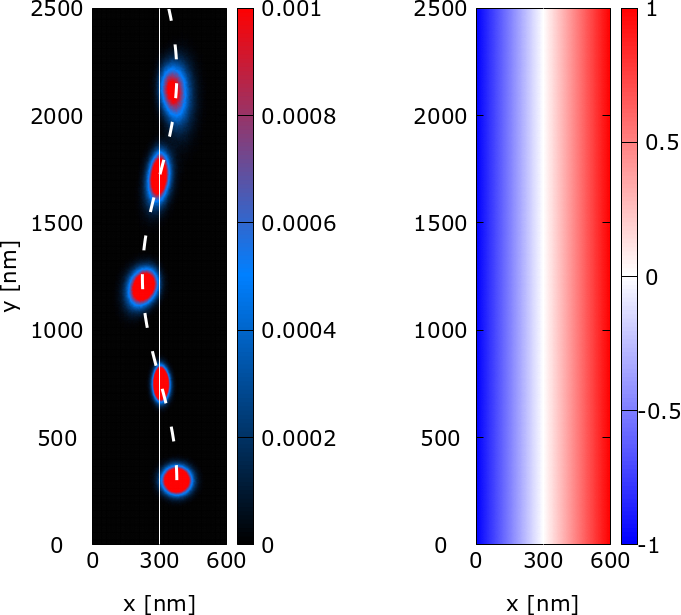}
\caption{Snake trajectory: Evolution of a minimum uncertainty Wigner state along a waveguide under an orthogonal, non-uniform linear magnetic field $B(x)=B_0 + B_1 x$ with $(B_0, B_1)=(-1\, \textrm{T}, 0.0033\, \textrm{T}\textrm{nm}^{-1} )$. The right part shows the magnetic field $B(x)$ in Tesla (T). The left part shows the electron density at $t=\{0, 1.6 \textrm{ps}, 3.2 \textrm{ps}, 5\textrm{ps}, 6.4\textrm{ps}\}.$ The electron state follows a snake trajectory indicated by the white dashed line.}
\label{fig:snake}
\end{figure}


\vspace*{2cm}

\begin{figure}[htbp]
\includegraphics[width=0.45\textwidth]{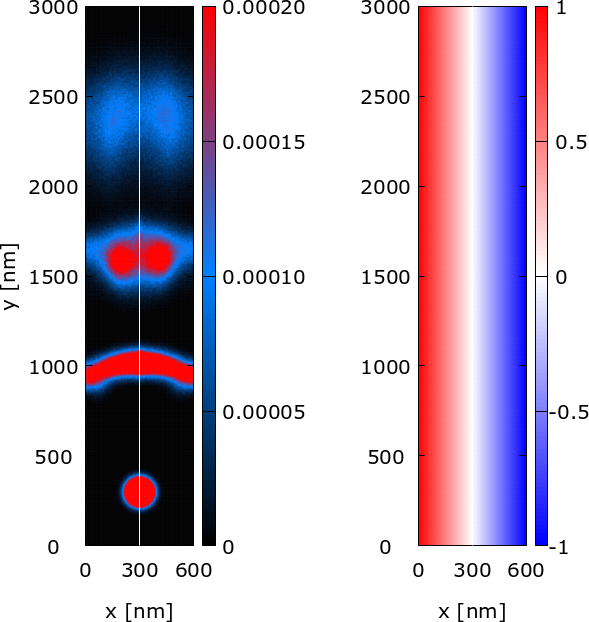}
\caption{Edge state: Evolution of a minimum uncertainty Wigner package along a waveguide under an orthogonal non-uniform linear magnetic field $B(x)=B_0 + B_1 x$ with $(B_0, B_1)=(1\, \textrm{T}, -0.0033\, \textrm{T}\textrm{nm}^{-1} )$. The right part shows the magnetic field $B(x)$ in Tesla (T). The left part shows the electron density at $t=\{0, 2.5 \textrm{ps}, 5\textrm{ps}, 8\textrm{ps}\}.$ At the end, the wavepacket becomes a split edge state.}
\label{fig:edge}
\end{figure}

\subsection{Snake Trajectory and Edge State}\label{sec_snake_edge}

In this section, we are going to explore the ability of a non-uniform magnetic field to control the evolution of an electron-state, not only in terms of trajectory and interference patterns, as in the magnetotunneling application, but also in terms of shape and dispersion of the wave-packet.
In particular, we are identifying two opposite configurations of the linear magnetic field where the first contributes to maintain the localization of the wave-packet, as it happens in the case of a snake trajectory, and the second to de-localize, which gives rise to the creation of an edge state. 

\subsubsection{Simulation Setup}
In the two following experiments, we consider a portion of an electronic waveguide in the plane $x-y$, with a magnetic field, $\Bv(\rv)$, that is orthogonal to the plane $x-y$ and presents a linear dependence in direction $x$, which is orthogonal to the electron evolution (electron state evolves along $y$). The waveguide width is $600\, \rm nm$ along $x$-direction. The magnetic field is defined by $\Bv(\rv)={(0,0,B(x))}$ with $B(x)=B_0+B_1 x$. Equation (\ref{eq:Wiglinmag}), that describes the electron state evolution, keeps the same form except that $y$ and $x$ are exchanged. 
The waveguide length for the snake trajectory and edge state experiment is $2500\, \rm nm$ and $3000\, \rm nm$, respectively. 
We evolve a minimum uncertainty Wigner state with $\sigma_{x} = \sigma_{y}=35\, \rm nm$, $m_\textrm{eff}=0.19 m_e$ and initial energy of $0.045 \, \rm eV$ along $y$.
The electron state is placed at $y_0 = 300 \rm nm$ and we use a Gaussian state that is fully included in the simulation domain at $t=0$. 

\subsection{Snake Trajectory} \label{snake_trajectory}

In the snake trajectory experiment the value of the magnetic field is defined by $B_0=-1\, \rm T$ and $B_1=0.0033\, \rm T nm^{-1}$. This means that the magnetic field starts at $-1\, \rm T$ at $x=0\, \rm nm$ and increases to $1\, \rm T$ at $x=600\, \rm nm$ and is constant along the $y$-direction, as shown in the right part of Fig.~\ref{fig:snake}. The initial position of the Wigner state is $(x_0,y_0)=(370\, \textrm{nm}, 300\, \textrm{nm})$ and is placed $70\, \rm nm$ to the right of the waveguide center which is at $x=300\, \rm nm$,  where $B=0\, \rm T$), see Fig.~\ref{fig:snake}. 

The left part of Fig.~\ref{fig:snake} shows the evolution of the electron state along the waveguide. 
Fig.~\ref{fig:snake} reports the density of the quantum states in specific time instants $t=\{ 0\, \textrm{ps}, 1.6\, \textrm{ps}, 3.2\, \textrm{ps}, 5.0\, \textrm{ps}, 6.4\, \textrm{ps}\}$. 
The dashed white line represents the trajectory of a classical "point-like" particle that evolves along the wire starting from the middle of the quantum state with the same mean velocity. 
Furthermore, the chosen magnetic field affects the shape of the electron state, as shown by the distribution of the electron density. At $t=0$, the state is perfectly circular. 
During the evolution the "width" of the wavepacket tends to decrease approaching the center ($t=1.6 \rm ps$), since the magnetic field decreases and reaches $0 \rm T$, followed by an increase again as it oscillates to the other side ($t=3.2 \rm ps$). 
On the contrary, the natural spreading (i.e., increase of the dispersion) characterising a freely evolving Gaussian state is retained in the direction of the evolution. 
This particular choice of the initial condition and of the external linear magnetic field allows to limit the dispersion of the electron state during the evolution and shows how a linear magnetic field can be used to control the shape of the density of an electron state during its evolution.

\subsection{Edge State}

In the edge state experiment, the magnetic force is opposite to the one used in Section~\ref{snake_trajectory}, as can be seen from the right part of Fig.~\ref{fig:edge}. Thus $B_0=1\, \rm T$ and $B_1=-0.0033\, \rm T nm^{-1}$, so the magnetic field ranges from $1\,\rm T$ to $-1\, \rm T$ along $x$, and is constant along the $y$-direction. The initial position of the Wigner state ($t=0\, \rm ps$) is perfectly centered, i.e., $(x_0,y_0)=(300\, \textrm{nm}, 300\, \textrm{nm})$, where $B=0\, \rm T$, as can be seen in the left part of Fig.~\ref{fig:edge}. 
In the classical case, a particle evolving in the center and along the line $x=300\,\rm nm$ (where the magnetic force is zero) will not be affected by the action of the magnetic field. 
In the quantum case, the electron state density is distributed in space, according to Heisenberg's uncertainty principle, and represented by an ensemble of numerical particles, so that the evolution is always affected by the action of the magnetic field.
It is thus clear that in the quantum case the signed-particle model is useful as it allows to describe and heuristically understand the electron evolution.
Parts of the electron density on the left of the central line are pushed towards the left border of the waveguide, while other parts on the right are pushed towards the right border. The boundary conditions reflect the parts of the split state back towards the center. 
An analysis of the left part of Fig.~\ref{fig:edge} shows the density of the quantum states at specific times $t=\{ 0 \, \textrm{ps}, 2.5 \, \textrm{ps}, 5.0 \, \textrm{ps}, 8.0 \, \textrm{ps} \}$.  As we can see at $t=2.5\, \rm ps$ the electron density width increases along the $x$-direction during the evolution. This dispersion is driven by the Lorenz force and provides a curvilinear shape along $x$ to the wavepacket, similar to an "arc". When the electron state interacts with the lateral borders of the waveguide it is reflected towards the center, this can be seen by the fact that the bending of the curvilinear shape disappears near the borders.
At $t=5.0\, \rm ps$ the electron state is completely reflected and the density takes  an almost "bimodal distribution" form along $x$. It is almost separated in two parts, one on the left and one on the right of the line $x=300\, \rm nm$, where the magnetic field is zero. Only a small portion of the density is still in the center of the waveguide. The electron density at $t=8.0\, \rm ps$ keeps the bimodal distribution which is more defined, and evolves along the waveguide as two parallel parts of an edge state that travel in the same direction, due to the opposite sign of the magnetic field. 
This behaviour is further confirmed by the dispersion of the two parts along the direction of motion, that is clearly visible comparing to the density at $t=5.0\, \rm ps$ with the density at $t=8.0\, \rm ps$. 
This configuration of the magnetic field splits the state and pushes its parts towards the boundaries which reflect them back.
This mechanism of electron splitting can be an attractive option for research in electron quantum optics, where "beam splitter" concepts are a vital building block. 

\section{Summary}

Non-uniform magnetic fields offer unique capabilities to control single electron states. Here, we investigate a generalization of equation (25) proposed in~\cite{PRA} for non-linear electric fields. The obtained equation involves the conventional Wigner potential, the magnetic Lorentz force, that includes also linear magnetic fields, and higher-order terms that depend on the linear coefficient of $\Bv$. The higher-order terms can be neglected for small magnitudes which allows to perform numerical calculations in terms of the well-developed signed-particle model. 
Experiments on magnetotunneling  are shown and, in particular, the capability of linear, non-uniform magnetic fields to control an electron state trajectory. 
Moreover, the possibility to change the value of the magnetic field allows to specifically influence an interference pattern or an oscillatory behaviour.  
We also show the capability of linear, non-uniform magnetic fields to control the spatial dispersion of an electron state, or to split the electron state. 
It should be stressed that both cases of evolution have a classical analog of evolving initial Gaussian distributions of an ensemble of non-interacting electrons. Indeed, for the considered physical setup classical and quantum evolution rules are the same. However, in the former, the initial condition can be only non-negative, while in the latter also negative values are possible, characterising the Wigner quasi-distribution function. 

\vspace{2cm}


\section*{Conflicts of interest}
There are no conflicts to declare.

\section*{Acknowledgements}
The financial support by the Austrian Science Fund (FWF): P33609 and P37080 is gratefully acknowledged.
The computational results have been achieved using the Vienna Scientific Cluster (VSC).




\balance


\bibliography{main} 
\bibliographystyle{rsc} 

\end{document}